%% file: main.tex
\documentclass[letterpaper, journal, 10pt]{IEEEtran}

\input{header_traditional}

\usepackage{enumitem}

\begin{document}

\title{Exact model reduction for\\ discrete-time conditional quantum dynamics}

\author{Tommaso Grigoletto and Francesco Ticozzi
\thanks{T. Grigoletto and F. Ticozzi are with the Department of Information Engineering, University of Padova, Via Gradenigo 6, 35131 Padova, Italy. Emails: 
\href{mailto:tommaso.grigoletto@unipd.it}{\texttt{tommaso.grigoletto@unipd.it}},
\href{mailto:ticozzi@dei.unipd.it}{\texttt{ticozzi@dei.unipd.it}}.\\ Acknowledgments: The authors wish to thank Lorenza Viola, Clément Pellegrini, and Tristan Benoist for stimulating discussions on this work. The work of F. T. was supported by European Union through NextGenerationEU, within the National Center for HPC, Big Data and Quantum Computing under Projects CN00000013, CN 1, and Spoke 10.}}

\maketitle
\begin{abstract}
Leveraging an algebraic approach built on minimal realizations and conditional expectations in quantum probability, we propose a method to reduce the dimension of quantum filters in discrete-time, while maintaining the correct distributions on the measurement outcomes and the expectations of some relevant observable. The method is presented for general quantum systems whose dynamics depend on measurement outcomes, hinges on a system-theoretic observability analysis, and is tested on prototypical examples.
\end{abstract}


\thispagestyle{empty}
\pagestyle{empty}

\section{Introduction}
Quantum stochastic dynamical models are key to the theory and practice of real-time feedback in quantum experiments \cite{sayrin2011real}, and at the same time represent one of its key limitations. In fact, despite their successful implementation in (relatively) small systems, the real-time integration of filtering equations poses significant limits to the effective bandwidth of the control. For these reasons, various methods have been considered to reduce the computational burden of these models, including projection filters and approximate reductions based on adiabatic-type limits \cite{ cardona2020exponential,  ramadan2022exact}.
In this work, we construct stochastic quantum models in discrete time that are able to exactly reproduce the output and measurement statistics for a wide class of quantum systems, including discrete-time filtering and feedback equations \cite{bolognani_ticozzi,bouten2009discrete, amini2013feedback}, quantum trajectories \cite{benoist2019invariant}, quantum walks including measurements \cite{attal2012open}, \cite{apers2018simulation} and partially observed quantum Markov process \cite{barry2014quantum}, while allowing for imperfect measurements and/or un-monitored environments.

Our approach builds on the ideas of \cite{itoIdentifiabilityHiddenMarkov1992}, (see \cite{schonhuth2008simple} for a generalization to quantum walks), where stochastic analogues of reachable and non-observable spaces are used to characterize equivalent hidden-Markov processes. Exploiting the ideas developed in \cite{tac2023,tit2023} for {unconditional} dynamics, the method leverages algebraic quantum probability tools, in particular conditional expectations, to allow for {\em reduced conditional models} that generate the same measurement distribution, reproduce the average of observables of interest, and do so while maintaining the same structure and physical constraints as the original model. In general, existing reduction approaches cannot guarantee that the reduced evolution is completely positive (CP) and trace-preserving (TP) \cite{tokieda2023complete}. Therefore, we show that the algebraic framework introduced in \cite{tit2023} is general and can be adapted to many cases of interest. Comparing the reduced models with and without conditioning, we highlight some key differences between classical and quantum processes: in some cases quantum conditioning can further reduce the minimal realization, while in the classical case this is never possible. It is worth remarking that the same approach can be adapted to derive reduced models from known initial conditions of interest, i.e. exploiting reachable reductions. This problem is less natural for filtering dynamics and requires the introduction and use of modified algebraic structures that go beyond the scope of the present work \cite{tit2023}. 

In Section \ref{sec:problem_def} we introduce the class of models of interest. Note that some of the assumptions introduced in Section \ref{sec:problem_def} for simplicity can be relaxed. For example, one could consider the case of multiple measurement apparatuses and evolution maps, among which one can choose which one to measure or apply. Section \ref{sec:linear_reduction} introduces the non-observable subspace and the optimal linear reduction. Similar ideas have been introduced, with different objectives in \cite{dalessandro_quantum_2003}. Section \ref{sec:quantum_reduction} illustrates how to extend a minimal observable representation of the dynamics to a CPTP one. Section \ref{sec:examples} is devoted to examples.
\section{Model and problem definition}
\label{sec:problem_def}
In this paper, we consider finite-dimensional Hilbert spaces $\Hc\simeq\Cb^n$. The algebra of linear operators (closed with respect to the standard matrix product) acting on $\Hc$ is denoted by $\Bf(\Hc)\simeq\Cb^{n\times n}$, and the set of density operators by $\Df(\Hc) = \{\rho|\rho\geq0, \, \tr(\rho)=1\}$.

\subsection{Dynamics of interest}

Let us assume that we are able to perform a generalized measurement \cite{kraus1983states} with outcomes $k\in\Omega$ described by the operators $\{M_k\}$ such that $\sum_k M_k^\dag M_k =\one$. Some of the scenarios of interest for this work include:\\ \noindent
$\bullet$ \textit{CPTP dynamics followed by generalized measurement}: assume that the system evolves through a CPTP dynamical map $\Ec(\cdot)$ (e.g. evolution in contact with a Markov bath) and after each evolution the generalized measurement is performed. In this case, the effects of conditioning, given the measurement outcome $k$, are described by a set of CP maps $\{M_k \Ec(\cdot) M_k^\dag\}$.\\\noindent
$\bullet$ \textit{Imperfect measurement} \cite{amini2013feedback}: Assume now that the measurement apparatus correctly registers the measurement outcome $k$ with (known) probability $p_{k,k},$ while with probability $p_{k,j}$ the actual outcome was $j,$ with $\sum_j p_{k,j} = 1$. Then the conditioning of the state given the registered outcome $k$ is given by a set of CP maps $\{\sum_j p_{k,j} M_j \cdot M_j^\dag \}$. \\\noindent
$\bullet$ \textit{Dynamics conditioned on the measurement outcome}: Assume to have a set of CPTP $\{\Ec_k\}$, one per outcome in $\Omega$ and assume that, after observing the measurement outcome $k$, the dynamical map $\Ec_k$ is applied (e.g. feedback unitary operations \cite{bolognani_ticozzi}). In this case, the effects of conditioning are described by a set of CP maps $\{\Ec_k(M_k\cdot M_k^\dag)\}$.

In the following, we propose a general framework capable of describing the presented cases in a compact form: \textit{conditional dynamics}.

Let us consider a quantum system defined over a Hilbert space $\Hc$. Let us assume to be able to perform a measurement on the system at hand (either projective, generalized, or imperfect). Let $\Omega$ denote the set of measurement outcomes and let $\Pb_\rho[\Mc=k]$ be the probability of observing the outcome $k\in\Omega$ when performing the measurement on the system in the state $\rho$.   
A \textit{quantum instrument} \cite{kraus1983states} is a set of CP super operators $\{\Mc_k\}_{k\in\Omega}$, one per every outcome in $\Omega$, such that $\sum_{k\in\Omega} \Mc_k^\dag(\one) = \one$. 
Assume that the system is prepared in a state $\rho$. The outcome $k\in\Omega$, is observed with probability 
\begin{equation}
    \mathbb{P}_\rho[\Mc=k] = \tr[\Mc_k (\rho)].    
\end{equation}
After the observation of the outcome, the system state needs to be conditioned on the observed outcome $k$, resulting in the state:
\begin{equation}
    \rho|_{\Mc=k} = \frac{\Mc_k(\rho)}{\tr[\Mc_k(\rho)]}.
    \label{eq:measurement_collapse}
\end{equation}

Notice that the normalization factor at the denominator of the state update \eqref{eq:measurement_collapse} makes the dynamics of the process non-linear. Linear dynamics can be recovered by considering the un-normalized state $\Tilde{\rho}|_{\Mc=k} = \Mc_k(\rho).$   We can further notice that the probability of the outcome coincides with the trace of the un-normalized state, i.e. $\Pb_\rho[\Mc = k] = \tr[\Tilde{\rho}|_{\Mc=k}]$, and that the normalized state $\rho|_{\Mc=k}$ can still be retrieved by \textit{a posterior} re-normalization, i.e. 
\(\rho|_{\Mc=k} = {\Tilde{\rho}|_{\Mc = k}}/{\tr[\Mc_k(\rho)]}.\)

\subsection{Quantum Conditional Evolutions}
\label{sec:model}
We now introduce a class of discrete-time quantum dynamical models that includes the effects of conditioning. In the following, we assume:
\textbf{1)} To have a quantum instrument, described by a set of CP maps $\{\Mc_k\}_{k\in\Omega}$ such that $\sum_{k\in\Omega} \Mc_k^{\dag}(\one) = \one$, where $\Omega$ is the set of possible outcomes; 
\textbf{2)} That a measurement is performed at each time step $t = 0, 1, 2,\dots$. Therefore, we will indicate the outcome observed at time $t$ with the subscript $t$, i.e. $\Mc(t)=k_t$ indicates the measurement at time $t$ leads to the outcome $k_t\in\Omega$. Moreover, we denote an ordered sequence of conditional dynamics as $\Mc_{k_{0:t}}$ and of outcomes as $k_{0:t} := k_0, k_1, \dots, k_k$ and thus \[\Pb_{\rho}[\Mc({0:t})=k_{0:t}] := \Pb_\rho[\Mc(0)=k_0, \dots, \Mc(t)=k_t].\]
\textbf{3)} We further assume to only be interested in reproducing the expectation value of a set of observables $\{O_j\}$ containing the identity, i.e. $\one\in\{O_j\}$. 
We compactly represent the set of observables of interest $\{O_j\}$ via a {\em linear output map} $\Cc: \Bf(\Hc)\to \Ys$ where $\Ys$ is a vector space of dimension $\dim(\Ys) =|\{O_j\}|$. This map is constructed by fixing a basis $\{E_j\}$ for $\Ys$ and imposing $\Cc(\cdot) = \sum_i E_i \tr[O_j \cdot]$.

In the rest of the manuscript, we write $\Tilde{\rho}(t)$ to denote $\Tilde{\rho}(t,\Mc({0:t-1})=k_{0:t-1})$, leaving the dependence on the outcomes implicit when the meaning of the symbols is clear from the context.
We shall consider a sequence of repeated quantum instruments, a \textit{conditional dynamics}, for the un-normalized density, described by 
\begin{equation}
    \Tilde{\rho}(t+1) = \Mc_{k_{t+1}}\left[\Tilde{\rho}(t)\right]
    \label{eq:evolution}
\end{equation}
so that $\Tilde{\rho}(t+1) = \Mc_{k_{0:t+1}}(\rho_0)$ where \[\Mc_{k_{0:t+1}} := \Mc_{k_t+1} \circ \Mc_{k_{t}} \circ \dots \circ \Mc_{k_1} \circ \Mc_{k_0}. \]
Notice that the evolution described by equation \eqref{eq:evolution} is  that of a switching model with linear dynamics, where the switching action depends on the outcome of the measurement $k_t\in\Omega$.

As noted before, the un-normalized state also allows for direct computation of the joint probabilities of trajectories, as \(\Pb_{\rho_0}[\Mc(0:t)=k_{0:t}] = \tr[\Tilde{\rho}(t+1,\Mc({0:t})=k_{0:t})]\)
and the corresponding actual density operator at any time $t$ can be computed by re-normalization with the probability of the trajectory:
\begin{align*}
    \rho(t+1| \Mc(0:t)=k_{0:t}) = \frac{\Tilde{\rho}(t+1, \Mc({0:t})=k_{0:t})}{\Pb_{\rho_0}[ \Mc({0:t})=k_{0:t}]}.
\end{align*}
In exactly the same manner, it is possible to retrieve the evolution of the expectation value of an observable $O$ conditioned on the outcomes $k_{0:t}$, i.e. $\expect{O(t+1|\Mc(0:t)=k_{0:t})}$, starting from the un-normalized state via \textit{a posterior} re-normalization, 
\(\tr[O \rho(t+1|\Mc(0:t)=k_{0:t}))] = {\tr[O \Tilde{\rho}(t+1,\Mc(0:t)=k_{0:t})) ]}/{\Pb_{\rho_0}[\Mc({0:t}) = k_{0:t}]}. \)
Thus, in order to reproduce the evolution of the expectation value of an observable $O$ we simply need to keep track of the evolution of the two quantities $\tr[O \Tilde{\rho}(t)]$ and $\tr[\Tilde{\rho}(t)]$.

The dynamic of interest in this work is captured in the following definition.
\begin{definition} 
\label{def:CQHM_model}
Let us consider a set of conditional dynamic $\{\Mc_k\}_{k\in\Omega}$ and let $\Cc$ be a linear output map.
We define a quantum {\em conditional evolution} (CE) as:
 \begin{equation}
    \begin{cases}
    \Tilde{\rho}(t+1) = \Mc_{k_{t}}[\Tilde{\rho}(t)]\\
    y(t) = \Cc\left[\Tilde{\rho}(t)\right]
    \end{cases}\quad \tilde\rho(0)=\rho(0)\in\Df(\Hc)
    \label{eqn:multi_time_model}
\end{equation}
We compactly denote a CE as the couple $(\{\Mc_k\},\Cc)$.
\end{definition}

With this definition, we can formally define the problem that is going to be tackled in this paper.  

\begin{problem}
\label{prob:multi_time}
Given a model as in \eqref{eqn:multi_time_model}, find a linear map $\Phi$ and another CE $(\{\check{\Mc}_k\},\check{\Cc})$,  of smaller dimension, and with initial condition $\Tilde{\tau}(0)=\Phi[\rho(0)]$
such that, for every initial state $\rho(0)\in\Df(\Hc)$ and every sequence of outcomes $k_{0:t}$, the two models provide the same outputs conditioned on the measured outcomes $k_{0:t}$ i.e. 
\(\Cc[\Tilde{\rho}(t+1, \Mc_{0:t}=k_{0:t})] = {\check{\Cc}}[{\Tilde{\tau}}(t+1, \Mc_{0:t}=k_{0:t})].\)
\end{problem}

\begin{remark}
Notice that, because we assumed that $\one\in\{O_i\}$ the above problem ensures that the reduced model is also capable of reproducing the probabilities of sequences of outcomes:
    \begin{align*}
    \Pb_{\rho_0}[\Mc({0:t})=k_{0:t}] &= \Pb_{\Phi(\rho_0)}[\check{\Mc}({0:t})=k_{0:t}]\\
    \tr[\Tilde{\rho}(t+1)] &= \tr[{\Tilde{\tau}}(t+1)].
\end{align*}
This can be considered as the zero-th order problem where the only observable we are interested in is the identity and we thus aim to find a reduced CE capable of reproducing the probability of the trajectories.
\end{remark}
\section{Observability and Linear Reduction}
\label{sec:linear_reduction}
In this section, we will extend the work of \cite{itoIdentifiabilityHiddenMarkov1992} to the quantum framework. We shall start by defining the minimal subspace that supports the dynamics. This is done by adapting well-known results of control system theory \cite{wonham} to the case of switching systems. 

Intuitively, to reduce the model, we can remove from the model description the states associated with indistinguishable states, i.e. states that provide identical output trajectories. These states are characterized through the non-observable subspace.
\begin{definition}[Non-observable subspace]
    Given a CE we define the non-observable subspace as
        \begin{align*}
            \Ns & = \left\{X\in\Bf(\Hc)|\quad\Cc[X]=0 \right.\\
                &\qquad\left.\text{ and } \Cc \left[\Mc_{k_{0:t}}[X]\right]=0, \, \forall k_{0:t}, t\geq0\right\}
            \label{eqn:cond_non_obs_def}
        \end{align*}
\end{definition}\noindent
A few key properties of this space are given.
\begin{proposition}
\label{prop:subspace_invarinace}
    $\Ns$ is $\Mc_{k}$-invariant for all $k\in\Omega$, and it is contained in $\ker\Cc$.
\end{proposition}
\begin{proof}
    From the definition, we have that $X\in\Ns$ if $\Cc[\Mc_{k_{0:l}}[X]]=0$ for any $l>0$. Then, for any $k\in\Omega$ we have that $\Mc_{k}[X]$ satisfies $ \Cc\Mc_{k_{0:l}}\Mc_k[X] = \Cc\Mc_{k_{0:l},k}[X]=0$, and thus $\Mc_k[X]\in\Ns$. From the definition of $\Ns$, we have that $\Ns\subseteq\ker\Cc$.
\end{proof}

Notice that the orthogonal subspace to $\Ns$ {(w.r.t the standard Hilbert-Schmidt inner product $\inner{\cdot}{\cdot}_{HS}$)} takes the form 
\begin{equation}
   \Ns^\perp = \Span\left\{O_j, \Mc_{k_{0:t}}^{\dag}(O_j),\, \forall k_{0:t}, \, t\geq0, \, \forall j \right\}. 
   \label{eqn:observable_subspace}
\end{equation}
{This can be derived from the definition by writing $\Cc(\Mc_{k_{0:t}}(X)) = \sum_i E_i \tr[O_j \Mc_{k_{0:t}}(X)],$  interpreting the trace as the HS inner product and switching to the dual dynamics $\Mc_{k_{0:t}}^\dag(O_j).$ It is then sufficient to notice that the latter operators are orthogonal to all the $X\in\Ns$.}
Moreover, one can also extend the results of \cite[Lemma 3]{itoIdentifiabilityHiddenMarkov1992} to prove that the sequences $k_{0:t}$ can be limited to sequences of length less or equal to $\dim(\Hc)^2$. Finally, we can notice that, because we assumed that $\one\in\{O_j\}$, we have that $\one\in\Ns^\perp$.
\vspace{-1em}
\subsection{Linear model reduction}
If one is not interested in having a quantum (CPTP) reduced model, the model can be reduced simply using the projector $\Pi_{\Ns^\perp}$ obtaining a linear model that reproduces the conditioned output of the quantum system.
The following Proposition formalizes and extends this fact.
\begin{proposition}
    Let $\Vs$ be an operator subspace that contains $\Ns^\perp$, i.e. $\Ns^\perp\subseteq\Vs$ and let $\Pi_\Vs$ be a projector onto $\Vs$. Then we have
    \[\Cc \Mc_{k_{0:t}}[\rho(0)] = \Cc \Pi_\Vs \Mc_{k_t}\Pi_\Vs\dots\Pi_\Vs \Mc_{k_0}\Pi_\Vs[\rho(0)]\]
    for all sequences $k_{0:t}$ and for all $\rho(0)\in\Df(\Hc)$.
    \label{thm:switching_reduction}
\end{proposition}
The proof of this follows from \cite[Theorem 4]{tac2023} {on the reducibility of switching dynamics. This can be applied to Proposition \ref{thm:switching_reduction} by vectorizing the state and obtaining the corresponding matrix form of the superoperators.}
To obtain the linear reduced model one can then find two factors of $\Pi_{\Ns^\perp}$, $\Rc_L:\Bf(\Hc)\to\Cb^q$ and $\Jc_L:\Cb^{q}\to\Ns^\perp$ with $q=\dim(\Ns^\perp)$, such that $\Pi_{\Ns^\perp} = \Jc\Rc$ and $\Rc\Jc = \one_q$. With these, one can construct a set of $q\times q$ matrices $\{A_k = \Rc_L \Mc_k \Jc_L\}$ and an output matrix $C=\Cc\Jc_L$ such that the reduced model 
\begin{align*}
    \begin{cases}
        x(t+1) = A_k x(t)\\
        y(t) = C x(t),
    \end{cases}
    x(0) = \Rc_L(\rho_0)\in\Cb^q,
\end{align*}
reproduces the outputs $y(t)$ conditioned on the measurement outcomes $k_{0:t}$.
Furthermore, by using known results from control system theory, such a model is observable and thus minimal.
This result shows that the complex behaviors of quantum multi-time probabilities can still be simulated with a simpler linear model. This aspect might be particularly useful for building minimal filters that can estimate the system state on a classical computer and hence reduce the computational overhead associated with feedback controls. 

Note that the fact that the reduction onto any subspaces containing $\Ns^\perp$ provides the correct conditioned output, as stated in proposition \ref{thm:switching_reduction}, plays a central role in the following section. 
\section{Quantum model reduction}
\label{sec:quantum_reduction}
\subsection{$*$-Algebras and conditional expectations}
We here briefly review the main results from quantum probability theory, finite-dimensional operator algebras, and conditional expectations that are necessary for this work. More details on the matter can be found in \cite{tit2023, petz2007quantum, wolf2012quantum}.

We define a finite-dimensional \textit{$*$-algebra} $\As\subseteq\Bf(\Hc)$ as an operator subspace closed with respect to the standard matrix product and the adjoint operator $\dag$ (transposed and complex conjugate), i.e. for all $X, Y\in\As$ and $\alpha,\beta\in\Cb$ $\alpha X + \beta Y\in\As$, $X^\dag,Y^\dag\in\As$ and $XY\in\As$. An algebra $\As$ is said to be \textit{unital} if it contains the identity $\one\in\As$.

Given an unital algebra $\As$ there exists a decomposition of the Hilbert space $\Hc = \bigoplus_k \Hc_{S,k} \otimes \Hc_{F,k}$ and a unitary change of basis $U\in\Bf(\Hc)$ that puts the algebra into a block-diagonal structure (called Wedderburn decomposition) of the form
\[U^\dag \As U = \bigoplus_k \Bf(\Hc_{S,k})\otimes \one_{F,k}.\]
An algebra with such a structure is thus isomorphic to an algebra $\check{\As}=\bigoplus_k\Bf(\Hc_{S,k}),$ which admits representation in a smaller Hilbert space.
Furthermore, any unital algebra admits a CPTP and unital orthogonal projector $\CE_\As:\Bf(\Hc)\to\As$, i.e. $\CE_\As^\dag = \CE_\As$, $\CE_\As(\one)=\one$ {(where the adjoint here is taken w.r.t the inner product $\inner{\cdot}{\cdot}_{HS}$)}. Such a map is a \textit{conditional expectation} (that preserves the completely mixed state $\one/n$) \cite{petz2007quantum}. Conditional expectations can be factorized into two CP factors, which we denote with $\Rc:\Bf(\Hc)\to\check{\As}$ and $\Jc:\check{\As}\to\As$ such that $\Rc\Jc = \Ic_{\check{\As}}$ the identity super operator over the algebra $\check{\As}$, and  $\Jc\Rc = \CE_\As$. Given the Wedderburn decomposition of $\As$, the conditional expectation and its CPTP factorization can be easily constructed \cite{wolf2012quantum,tit2023}. In the case of orthogonal conditional expectations $\CE_\As$ gets factorized as $\CE_\As=\Jc\circ\Rc$ where:
\begin{align}
    \label{eqn:reduction}
    \Rc(X)&=\bigoplus_k \tr_{\Hc_{F,k}}(W_k X W_k^\dag)=\bigoplus_k X_{S,k}=\check X\\
    \label{eqn:injection}
    \es(\check X)&=U\left(\bigoplus_k X_{S,k} \otimes\one_{F,k}/\dim(\Hc_{F,k})\right)U^\dag.
\end{align}
where we defined isometries $W_k^\dag:\Hc_{S,k}\otimes\Hc_{F,k}\to\Hc$.
\vspace{-1em}
\subsection{Quantum reduction}
We can define an observable algebra for the conditional system as follows. 
\begin{definition}[Output algebra]
    Let us consider a CE and let $\Ns$ be its non-observable subspace. Then we name $\As = \alg(\Ns^{\perp})$ the \textit{output algebra}.
\end{definition}

We are now ready to introduce the proposed solution of Problem \ref{prob:multi_time} which is compactly summarized in Algorithm \ref{algo:observable_mutli_time}.

\begin{algorithm}
    \caption{Projection onto the output algebra.}
    \label{algo:observable_mutli_time}
    \SetAlgoLined
    \Input{A {CE} $(\Mc_k, \Cc)$.}
    Compute $\Ns^\perp$\;
    Compute the output algebra $\As=\alg(\Ns^\perp)$\;
    Compute the factorization of $\CE_\As$, $\rs$ and $\es$\;
    \Output{$(\{\check{\Mc}_k = \Rc\Mc_k\Jc\}, \check{\Cc}=\Cc\Jc)$, $\Phi=\Rc$}
\end{algorithm}
We next prove that Algorithm \ref{algo:observable_mutli_time} solves Problem \ref{prob:multi_time}.
\begin{proposition}
\label{prop:reduction_multi_time}
    Let us consider a CE $(\{\Mc_k\}, \Cc)$ with observation algebra $\As=\alg(\Ns^\perp)$. Let $\CE_\As$ be the conditional expectation onto $\As$ and $\Rc$ and $\Jc$ its CPTP factorization. Then, fixing $\Phi=\rs$, the reduced CE defined over $\check{\As} := {\rm Im}(\rs)$, of conditional dynamics ${\check{\Mc}}_k := \rs\Mc_k\es$, output map $\check{\Cc} := \Cc\es$ and initial condition $\tau(0)=\Rc[\rho_0]$ solves Problem \ref{prob:multi_time}, that is for all sequences $k_{0:t}$ and for all $t\geq 0$ and $\rho_0\in\Df(\Hc)$ 
    \begin{align*}
       \Cc\Mc_{k_{0:t}}[\rho_0] = \check{\Cc}\check{\Mc}_{k_{0:t}}\Rc[\rho_0].
    \end{align*}
\end{proposition}
\begin{proof}
    Let us start by noticing that $\As$ is unital since $\one\in\Ns^\perp$. This implies that $\As$ allows for a CPTP conditional expectation $\CE_\As$ that can be factorized into two CPTP maps $\CE_\As = \es\rs$. Since the maps $\es$ and $\rs$ are CPTP, $\Rc[\rho_0]$ is a set of density operators, and $\{{\check{\Mc}}_k\}$ is a set of CP maps such that $\sum_{k\in\Omega}\check{\Mc}_k^\dag(\one) = \sum_k \es^\dag \Mc_k^\dag \rs^\dag (\one) = \one$, since $\rs^\dag$ and $\es^\dag$ are unital. 
    Recalling then that $\rs\es = \Ic_{\check{\As}}$ the identity super operator over $\check{\As}$ and $\CE_\As = \es\rs$ the CPTP projector onto $\As$, we have that \[\check{\Cc}{\check{\Mc}}_{k_{0:t}}\rs[\rho_0] = \Cc\left[ \CE_\As \Mc_{k_t}\CE_\As \dots \CE_\As\Mc_{k_0}\CE_\As (\rho_0)\right] \] 
    for all $\rho_0\in\Df(\Hc)$ and sequences $k_{0:t}$ with $t\geq0$. By noticing that $\Ns^\perp\subseteq\As$ and using Proposition \ref{thm:switching_reduction} we can conclude the proof.
\end{proof}

Using the same arguments as in \cite{tit2023}, one can prove that the closure to an algebra of $\Ns^\perp$, orthogonal to $\Ns$ with respect to the Hilbert-Schmidt inner product $\inner{\cdot}{\cdot}_{HS}$, provides the smallest algebra containing $\Ns^\perp$ that admits a CPTP projection, even when one allows for alternative definitions of operator product and orthogonality notions. 

\subsection{Reduction of measurements and dynamics, separately}
In the previous sections, we presented how to reduce a CE to a reduced-order one, obtaining a set of conditional dynamics that are capable of reproducing the output of interest conditioned on the outcome of the measurement. In many cases of interest, the conditional dynamic emerges as the composition of an unconditional CPTP evolution and a generalized measurement \cite{amini2013feedback}. Under suitable assumptions, it is possible to obtain reduced models for both the CPTP dynamical map and the measurements, separately. In the following we thus {\em assume that the conditional dynamics $\Mc_k$ are composed of a CPTP evolution map $\Ec$ following the conditioning induced by the generalized measurement} $\Kc_k(\cdot) := M_k \cdot M_k^\dag$ with $\sum_{k\in\Omega}M_k^\dag M_k = \one$, i.e. $\Mc_k(\cdot)= \Ec\circ\Kc_k$. This ordering choice has been made to keep consistency with previous works \cite{itoIdentifiabilityHiddenMarkov1992,tac2023} but similar results can be derived if the evolution precedes the measurement. 

\begin{assumption} Assume one of the following:
\begin{enumerate}[label=\textbf{(A\arabic*)}, nosep, wide, labelwidth=!, labelindent=0pt]
    \item $\exists\{\lambda_{k}\}$ such that $\sum_{k}\lambda_{k}\Mc_{k}(\cdot) = \Ec(\cdot)$\label{ass:1};
    \item The subspace $\Ns$ is $\Ec$-invariant\label{ass:2};
    \item The observation algebra $\As$ is $\Kc_{k}$-invariant for all $k$ \label{ass:3};
    \item The observation algebra $\As$ is $\Ec^{\dag}$-invariant \label{ass:4}.
\end{enumerate}

\end{assumption}
Assumption \ref{ass:1} is satisfied in two natural cases: \textbf{1)} when we have non-zero probability of skipping the measurement, i.e. $\lambda\one\in\{M_k\}$ for some $\lambda\in(0,1)$; \textbf{2)} when the model is defined on a commutative algebra $\Cs$, e.g. Hidden Markov Models, where we know that $\sum_k M_k\cdot M_k^\dag = \Ic_\Cs(\cdot)$. 

Notice that Assumption \ref{ass:1} implies Assumption \ref{ass:2}: by definition $\Ns$ is $\Mc_k$-invariant for all $k$, hence for $X\in\Ns$, we have $\Ec[X] = \sum_k \lambda_k \Mc_k[X]\in\Ns$. 
This generalizes the results presented in \cite[equation 3.3]{itoIdentifiabilityHiddenMarkov1992} and \cite[Lemma 4]{tac2023} to the case of stochastic output matrices (not only deterministic ones). Assumption \ref{ass:2} however, does not imply  \ref{ass:1}.

\begin{proposition}
\label{prop:separability}
    Under any of the assumptions \ref{ass:1}-\ref{ass:4}, the reduced CE $(\{\check{\Mc}_k\},\check{\Cc})$ with conditional dynamics $\check{\Mc}_k = \check{\Ec}\check{\Kc}_k$ where $\check{\Kc}_k = \Rc\Kc_k\Jc$ and $\check{\Ec} = \Rc\Ec\Jc$, $\check{\Cc} = \Cc\Jc$ and  $\Phi=\Rc$ solves Problem \ref{prob:multi_time}.
\end{proposition}
\begin{proof}
    We shall start by proving that \ref{ass:2} implies \(\Pi_{\Ns^\perp} \CE_\As \Ec \Kc_k \CE_\As = \Pi_{\Ns^\perp} \CE_\As \Ec \CE_\As \Kc_k \CE_\As \). From the fact that $\Ns$ is both $\Mc_k$- and $\Ec$-invariant we have $\Pi_{\Ns^\perp}\Mc_k = \Pi_{\Ns^\perp}\Mc_k\Pi_{\Ns^\perp}$ for all $k$ and $\Pi_{\Ns^\perp}\Ec = \Pi_{\Ns^\perp}\Ec\Pi_{\Ns^\perp}$. Moreover, from the fact that $\Ns^\perp\subseteq\As$ we have $\CE_\As\Pi_{\Ns^\perp} = \Pi_{\Ns^\perp}$. Substituting these relations into the left and right side of the statement we have, respectively $\Pi_{\Ns^\perp}\CE_\As\Ec\Kc_k\CE_\As = \Pi_{\Ns^\perp}\Ec\Kc_k\Pi_{\Ns^\perp}$ and $\Pi_{\Ns^\perp}\CE_\As \Ec \CE_\As \Kc_k  \CE_\As = \Pi_{\Ns^\perp}\Ec\Kc_k\Pi_{\Ns^\perp}$.

    We can then observe that under Assumption \ref{ass:3} we have $\Kc_k\CE_\As = \CE_\As\Kc_k\CE_\As$, while, under Assumption \ref{ass:4} we have $\CE_\As\Ec = \CE_\As\Ec\CE_\As$. By simple substitutions, under these conditions one obtains \(\CE_\As\Ec\Kc_{k}\CE_\As = \CE_\As\Ec\CE_\As\Kc_{k}\CE_\As.\)

   This, together with the proof of Proposition \ref{prop:reduction_multi_time}, shows that the conditional dynamics of the reduced model can be reduced separately into the reduced conditioning effects and the reduced dynamical map. 
\end{proof}

\section{Applications}
\label{sec:examples}
\subsection{Measured quantum walks}

Let us consider a finite-dimensional quantum system defined over the algebra $\Bf(\Hc)=\Cb^{n\times n}$. Let $O$ be an observable with no degeneracies, i.e. $O=\sum_{j=0}^{n-1}o_j\ketbra{j}{j}$ and $o_j\neq o_k$ for all $j\neq k$. The projective measurement of $O$ can be modelled by the set of superoperators$\{\Kc_j(\cdot) = \ketbra{j}{j} \cdot \ketbra{j}{j} \}$. Let us then assume that the system evolves through a unitary evolution map $\Ec(\cdot) = U\cdot U^\dag$ with $U$ such that $\Span\{\Ec^t[\ketbra{j}{j}], \forall t\geq0\} = \Bf(\Hc)$, By standard controllability considerations, a generic $U$ will satisfy this condition \cite{dalessandro_quantum_2003}. This implies that there is no unconditional model reduction for this model \cite{tit2023}. 

On the contrary, if we consider the effects of conditioning, we have that  $\Ns^\perp = \Span\{\ketbra{j}{j}\},$ which is an abelian algebra. Moreover, we can observe that $\Ns^\perp$ is $\Kc_k$ invariant for all $k$ and hence Proposition \ref{prop:separability} holds. This leads to a reduced model defined over an abelian algebra of dimension $n$, which is equivalent to classical Markov model defined over $\Rb^n$ with evolution map $P\in\Rb^{n\times n}$ such that $[P]_{j,k} = |\bra{j}U\ket{k}|^2$ and $\bm{1}^T P = \bm{1}^T$ the vector of all ones, output map $C= I_n$ and reduction map $\rs(\cdot) = \sum_j \ketbra{j}{j} \cdot \ket{j} = \diag(\cdot)$. 

 Situation like this, in which conditioning induces {\em smaller} reduced models, are not possible in the classical case. In fact \cite{itoIdentifiabilityHiddenMarkov1992} proves that the conditional observable space always contains the unconditional one.
\vspace{-1em}
\subsection{Measured Ising spin chain}
{The next model has been inspired by recent work on continuously-measured spin chains, see e.g. \cite{tirrito_full_2023}.} Let us consider a quantum system composed of $N\geq4$ qubits disposed on a line\footnote{We denote the operator living in $\Bf(\Hc)$ and acting as the Pauli matrix $\sigma_j$ with $j\in\{0,x,y,z\}$ on the $k$-th spin, with $k=1,\dots,N$ and as the identity on the rest is denoted with $\sigma_j^{(k)}$, i.e. $\sigma_j^{(k)} := \otimes_{i=1}^{k-1} \sigma_0 \otimes \sigma_j \otimes_{j=1}^{N-k}\sigma_0\in\Bf(\Hc)$.}, i.e. $\Hc \simeq (\Cb^2)^{\otimes N}$. Let us also assume to be able to perform projective measurement of the observable $\sigma_z$ on the last spin, i.e. $\sigma_z^{(N)}$ and that, at each step, this measurement is performed with probability $1-p$ with $p\in [0,1]$ and it is left untouched with probability $p$. Under these assumptions we can model the measurement with three super operators $\Kc_{-1}(\cdot) = p\one \cdot \one$, $\Kc_{0}(\cdot) = (1-p)\Pi_0 \cdot \Pi_0$ with $\Pi_0 := \one_{2^{N-1}}\otimes\ketbra{0}{0}$ and $\Kc_1(\cdot) = (1-p)\Pi_1 \cdot \Pi_1$ with $\Pi_1=\one_{2^{N-1}}\otimes\ketbra{1}{1}$. 
Between two measurements the system evolves with a unitary evolution $\Ec(\cdot) = U \cdot U^\dag$ with $U = e^{-i H}$ where $H$ is an Ising Hamiltonian, i.e. \[H = \delta \sum_{j=1}^{N-1}  \sigma_x^{(j)}\sigma_x^{(j+1)}\] with coupling strength $\delta$.
For this example, we are interested in reproducing the reduced state on the first qubit, i.e. $\tau(t) = \tr_{\bar{1}}(\rho(t))$, for all initial conditions $\rho_0\in\Df(\Hc)$. {Here, $\tr_{\bar{1}}(\cdot)$ denotes the partial trace with respect to everything but the first qubit of the chain,} which can be reconstructed from the expectation of  $\left\{\sigma_q^{(1)},\,{q=0,x,y,z}\right\}$, representing our observables of interest.
We next derive the reduced model in two cases of interest: $p=0$ and $0<p<1$ and we denote the spaces and maps of interest with the index $0$ and $p$ to distinguish them. 

\underline{\textbf{Case $p=0$.}}
Assuming $\delta\neq\frac{\pi}{2}j+\frac{\pi}{4}$ for some $j\in\Nb$, on can prove that 
\begin{align*}
    \Ns^{\perp}_0 &= \Span\{\sigma_{q}^{(1)}, \sigma_y^{(1)}\sigma_x^{(2)}, \sigma_z^{(1)}\sigma_x^{(2)}, \sigma_{q}^{(1)}\sigma_{z}^{(N)},\\&\qquad\sigma_y^{(1)}\sigma_x^{(2)}\sigma_{z}^{(N)}, \sigma_z^{(1)}\sigma_x^{(2)}\sigma_{z}^{(N)}, \, \forall q\in\{0,x,y,z\}\}
\end{align*}
of dimension $\dim(\Ns_0^{\perp})=12.$ Let us then define $\Bs_0 := \Cb^{2\times 2}$,   $\Bs_1:=\Span\{\sigma_0,\sigma_x\}\simeq \Cb\oplus\Cb$ and $\Bs_2=\Span\{\sigma_0,\sigma_z\}\simeq\Cb\oplus\Cb$. One then finds, 
\[ \As_0:=\alg\left(\Ns^{\perp}_0\right) = \Bs_0\otimes\Bs_1\otimes I_{2^{N-3}}\otimes \Bs_2\]
which is isomorphic to $\alg\left(\Ns_0^{\perp}\right)\simeq \oplus_{k=0}^3\Cb^{2\times2} =:\check{\As}_0$ regardless of the number $N$ of spins in the chain.
One can then verify that $\alg\left(\Ns_0^{\perp}\right)$ is both $\Kc_{0}$- and $\Kc_{1}$-invariant, hence Assumption \ref{ass:3} holds and we can separate the reduced dynamics and reduced effects of conditioning. 
Let $W_0$ be a unitary matrix that provides the Wedderburn decomposition
of $\alg(\Ns_0^{\perp})$, i.e. such that $W_0^\dag \As_0 W_0 = \oplus_{k=0}^{3}(\Cb^{2\times2} \otimes \one_{2^{N-3}})$.
The reduction super operator is then given by 
\[\bigoplus_{k=0}^3 \omega_k(t) = \Rc_0(\rho(t)) = \bigoplus_{k=0}^3 \tr_{\Hc_{F,k}}\left[ V_k W_0^\dag \rho(t) W_0^\dag V_k^\dag\right]\]
with $V_k^\dag:\Hc_k\to\Hc$ isometries of the correct dimensions and where $\Hc_{k} = \Hc_{S,k}\otimes\Hc_{F,k} \simeq \Cb^2\otimes \Cb^{2^{N-3}}$. 
The reduced evolution map takes the form $\check{\Ec}_0(\cdot) = \check{U}_0\cdot \check{U}_0^\dag$ with \(\check{U}_0 = \exp(-i\delta[\sigma_z\otimes\sigma_0\otimes\sigma_x+\sigma_0\otimes\sigma_y\otimes\sigma_0])\) and the reduced state on the first qubit is retrieved by $\tau(t) = \sum_{k=0}^3\omega_k(t)$. Moreover, the effects of conditioning for the reduced model are given by $\check{\Kc_0}(\cdot) = \check{\Pi}_0\cdot\check{\Pi}_0$ and $\check{\Kc_1}(\cdot) = \check{\Pi}_1\cdot\check{\Pi}_1$ with $\check{\Pi}_0 = \one_2\oplus\zero_2\oplus\one_2\oplus \zero_2$ and $\check{\Pi}_1 = \zero_2\oplus\one_2\oplus\zero_2\oplus\one_2$.

\underline{\textbf{Case $p\in(0,1)$.}}
Once again, one can verify that 
\begin{align*}
    \Ns^{\perp}_p &= \Span\{\sigma_{q}^{(1)}, \sigma_y^{(1)}\sigma_x^{(2)}, \sigma_z^{(1)}\sigma_x^{(2)}, \sigma_{q}^{(1)}\sigma_{z}^{(N)}, \\
    &\qquad \sigma_y^{(1)}\sigma_x^{(2)}\sigma_{z}^{(N)}, \sigma_z^{(1)}\sigma_x^{(2)}\sigma_{z}^{(N)}, \\
    &\qquad \sigma_{q}^{(1)}\sigma_x^{(N-1)}\sigma_y^{(N)}, \sigma_y^{(1)}\sigma_x^{(2)}\sigma_x^{(N-1)}\sigma_y^{(N)},\\
    &\qquad\sigma_z^{(1)}\sigma_x^{(2)}\sigma_x^{(N-1)}\sigma_y^{(N)},
    \, \forall q\in\{0,x,y,z\}\}
\end{align*}
of dimension $\dim(\Ns_p^{\perp})=18.$ Defining
$\Bs_3=\Span\{\sigma_0\otimes\sigma_0,\sigma_0\otimes\sigma_z,\sigma_x\otimes\sigma_y,\sigma_x\otimes\sigma_x\}\simeq\Cb^{2\times 2}$ one finds, 
\[ \As_p := \alg\left(\Ns^{\perp}_p\right) = \Bs_0\otimes\Bs_1\otimes I_{2^{N-4}}\otimes \Bs_3\]
which is isomorphic to $\alg\left(\Ns_p^{\perp}\right)\simeq \Cb^{4\times4}\oplus\Cb^{4\times 4} =: \check{\As}_p$ regardless of the number $N$ of spins in the chain. In this case, both Assumptions \ref{ass:1} and \ref{ass:3} hold. Let $W_p$ be a unitary matrix that provides the Wedderburn decomposition
i.e. such that $W_p^\dag \As_p W_p = \oplus_{k=0}^{1}(\Cb^{4\times4} \otimes \one_{2^{N-3}})$.
The reduction super operator is then given by 
\[\bigoplus_{k=0}^1 \xi_k(t) = \Rc_p(\rho(t)) = \bigoplus_{k=0}^1 \tr_{\Hc_{F,k}}\left[ V_k W_p^\dag \rho(t) W_p^\dag V_k^\dag\right]\]
with $V_k^\dag:\Hc_k\to\Hc$ isometries of the correct dimensions and where $\Hc_{k} = \Hc_{S,k}\otimes\Hc_{F,k} \simeq \Cb^4\otimes \Cb^{2^{N-3}}$. The reduced evolution map takes the form $\check{\Mc}_p(\cdot) = \check{U}_p \cdot \check{U}_p^\dag$ where, {\[\check{U}_p = \exp\left(-i\delta \bigoplus_{k=0}^1[\sigma_0\otimes\sigma_z + (-1)^k \sigma_x\otimes\sigma_0 ] \right)\]} and the reduced state on the first qubit is retrieved by $\tau(t) = \sum_{k=0}^1\tr_{\bar{1}}[\xi_k(t)]$.
Moreover, conditioning for the reduced model is associated to $\check{\Kc_0}(\cdot) = p\one_8\cdot\one_8$, $\check{\Kc_0}(\cdot) = (1-p)\hat{\Pi}_0\cdot\check{\Pi}_0$ and $\check{\Kc_1}(\cdot) = (1-p) \check{\Pi}_1\cdot\check{\Pi}_1$ with $\check{\Pi}_0 = \frac{\one_8 + \one_4\otimes \sigma_x}{2}$ and $\check{\Pi}_1 = \frac{\one_8 - \one_4\otimes \sigma_x}{2}$.

{In both cases treated above, only the first two and last two qubits of a chain (of arbitrary length) influence the dynamics of the first one. Moreover, the relevant degrees of freedom in the second spin are classical, associated to the eigenbasis of $\sigma_x$. With this, we have that the useful information behaves like the first qubit $\Cb^{2\times 2}$ coupled classical probabilistic mixture of two qubits, i.e.  $\Cb^{2\times2}\oplus\Cb^{2\times2}$.} 
\section{Conclusions}

We developed an algebraic method to reduce the dimension of 
discrete-time quantum systems, subject to repeated (generalized, imperfect) measurement and conditional dynamics, when
only the expectations of certain observables have to be reproduced exactly. Possible applications of the results include: (1) obtaining more intuitive and easier-to-study models; (2) implementing more efficient simulations of quantum trajectories on both classical and quantum computers; 
(3) efficiently producing output samples emerging from quantum repeated-measurement models; (4) deriving reduced-state estimators to increase the capability of feedback controllers for quantum systems. 
Two applications models have been discussed in details, highlighting interesting features and distinguishing the quantum case from its classical analogues.

A similar approach can be developed to obtain reductions leveraging knowledge of initial conditions, extending the results of \cite{tit2023} to the present setting. Future developments will be devoted to extending these ideas to the continuous-time case, finding reduction bounds and symmetry-based interpretations of the obtained models, and applying the methods to concrete situations of interest.
\bibliography{ref}

\end{document}

%% file: header_traditional.tex
\usepackage{amsfonts,amsmath,amssymb}
\usepackage[english]{babel}
\usepackage[normalem]{ulem}
\usepackage{hyperref}

\usepackage{amsthm}
\usepackage{graphicx}
\usepackage{subcaption,caption}
\usepackage{epsfig}
\usepackage{ulem}
\usepackage{tikz}
\usepackage{mathrsfs}
\usepackage{bbold}             
\usepackage{bm}
\usetikzlibrary{shapes,arrows}
\usetikzlibrary{matrix}
\usetikzlibrary{calc,patterns,angles,quotes,arrows,automata}
\graphicspath{ {./imgs/}}

\usepackage[linesnumbered,noend]{algorithm2e}
\RestyleAlgo{ruled}
\SetKwInOut{Input}{Input}
\SetKwInOut{Output}{Output}
\SetKwInOut{Parameters}{Parameters}
\SetKwIF{If}{ElseIf}{Else}{if}{:}{else if}{else}{endif}

\theoremstyle{definition}

\usepackage{mathabx}
\renewcommand{\check}{\widecheck}
\renewcommand{\bar}{\overline}

\theoremstyle{theorem}
\newtheorem{problem}{Problem}

\newtheorem{proposition}{Proposition}

\newtheorem{definition}{Definition}
\newtheorem*{assumption}{Assumptions}
\theoremstyle{definition}

\theoremstyle{remark}
\newtheorem{remark}{Remark}

\usepackage[numbers,sort&compress]{natbib}
\input{commands}

%% file: commands.tex
\newcommand{\Cc}{\mathcal{C}}

\newcommand{\Ec}{\mathcal{E}}

\newcommand{\Hc}{\mathcal{H}}
\newcommand{\Jc}{\mathcal{J}}
\newcommand{\Rc}{\mathcal{R}}
\newcommand{\Mc}{\mathcal{M}}

\newcommand{\Ic}{\mathcal{I}}
\newcommand{\Kc}{\mathcal{K}}

\newcommand{\As}{\mathscr{A}}
\newcommand{\Bs}{\mathscr{B}}
\newcommand{\Cs}{\mathscr{C}}

\newcommand{\Ns}{\mathscr{N}}
\newcommand{\Vs}{\mathscr{V}}

\newcommand{\Ys}{\mathscr{Y}}

\newcommand{\Rb}{\mathbb{R}}
\newcommand{\Cb}{\mathbb{C}}
\newcommand{\Pb}{\mathbb{P}}

\newcommand{\Nb}{\mathbb{N}}
\newcommand{\Bf}{\mathfrak{B}}

\newcommand{\Df}{\mathfrak{D}}

\newcommand{\CE}{\mathbb{E}|}

\newcommand{\rs}{\mathcal{R}}
\newcommand{\es}{\mathcal{J}}

\newcommand{\one}{\mathbb{1}}
\newcommand{\zero}{\mathbb{0}}

\newcommand{\tr}{{\rm tr}}
\newcommand{\diag}{{\rm diag}}
\newcommand{\Span}{{\rm span}}

\newcommand{\alg}{{\rm alg}}

\newcommand{\ket}[1]{\left| #1 \right>}
\newcommand{\bra}[1]{\left< #1 \right|}

\newcommand{\inner}[2]{\left< #1,#2 \right>}

\newcommand{\expect}[1]{\left< #1 \right>}

\newcommand{\ketbra}[2]{\left\vert #1 \middle>\middle< #2 \right\vert}

\definecolor{darkviolet}{rgb}{0.58, 0.0, 0.83}
\definecolor{lavender}{rgb}{0.45, 0.31, 0.59}